\documentclass[letterpaper,titlepage,11pt]{article}
\usepackage{jheppub}
\usepackage{hyperref}
\usepackage{epsfig}
\usepackage{graphicx}
\usepackage{amsmath}
\usepackage{amsfonts}
\usepackage{amssymb}
\usepackage{mathrsfs}
\usepackage{fancybox}
\usepackage{ragged2e}
\usepackage{tikz}
\usepackage{wasysym}

\newcommand{\mycite}[1]{~{\cite{#1}}}
\newcommand{\myfigref}[1]{~{Fig.~(\ref{#1})}}
\newcommand{\myref}[1]{~{(\ref{#1})}}
\newcommand{\tr}{\mathop{{\rm Tr}}}
\newcommand{\beq}{\begin{equation}}
\newcommand{\eeq}{\end{equation}}

\begin{document}

\begin{titlepage}

\rightline{\vbox{\small\hbox{\tt } }}
 \vskip 1.8 cm

\centerline{\LARGE \bf Matching three-point functions } \vskip
0.3cm \centerline{\LARGE \bf of BMN operators at weak and strong
coupling} \vskip 1.5cm

\centerline{\large {\bf Gianluca Grignani$\,^{1}$},  {\bf A.~V.~Zayakin$\,^{1,2}$}}

\vskip 1.0cm

\begin{center}
\sl $^1$ Dipartimento di Fisica, Universit\`a di Perugia,\\
I.N.F.N. Sezione di Perugia,\\
Via Pascoli, I-06123 Perugia, Italy
\vskip 0.4cm
\sl $^2$ Institute of Theoretical and Experimental Physics,\\
B.~Cheremushkinskaya ul. 25, 117259 Moscow, Russia
\end{center}
\vskip 0.6cm

\centerline{\small\tt grignani@pg.infn.it, a.zayakin@gmail.com }

\vskip 1.3cm \centerline{\bf Abstract} \vskip 0.2cm \noindent
The agreement between string theory and field theory is
demonstrated in the leading order by providing the first
calculation of the correlator of three
two-impurity BMN states with all non-zero momenta. The calculation
is performed in two completely independent ways: in field theory
by using the large-$N$ perturbative expansion, up to the terms
subleading in finite-size, and in string theory by using the
Dobashi-Yoneya 3-string vertex in the leading order of the Penrose expansion.
The two results come out to be completely identical.

\end{titlepage}

\addtocontents{toc}{\protect\setcounter{tocdepth}{1}}

\setcounter{page}{1}
\section{Introduction and motivations}
A very fascinating progress in computing three-point functions for
$\mathcal{N}=4$ Super Yang-Mills (SYM) operators has taken place
during the last two years\mycite{Janik:2010gc,Buchbinder:2010vw,Costa:2010rz,Roiban:2010fe,
Zarembo:2010rr,Hernandez:2010tg,Arnaudov:2010kk,Georgiou:2010an,Escobedo:2010xs,Park:2010vs,Russo:2010bt,Bak:2011yy,
Bissi:2011dc, Arnaudov:2011wq,Hernandez:2011up,Bai:2011su,
Ahn:2011zg,Klose:2011rm,Arnaudov:2011ek,Michalcik:2011hh,
Janik:2011bd,
Escobedo:2011xw,Gromov:2011jh,Georgiou:2011qk,Buchbinder:2011jr,Foda:2011rr,
Bissi:2011ha,Georgiou:2012zj,Gromov:2012vu,Serban:2012dr,Kostov:2012jr}
The global aim of these efforts would be to provide the full set
of three-point correlators for arbitrary number of constituent
fields, number of colours $N$ and t'Hooft coupling $\lambda$.
Given the conformal invariance of $\mathcal{N}=4$ SYM, this would
be equivalent to a full solution of the theory. In fact, two point
correlation functions of local operators can be determined by the
anomalous dimensions of the gauge theory operators and all the
higher point correlation functions can be reconstructed using
OPE's with the three point function structure constants. This
ambitious project is far from having been completed. While we have
a complete understanding of the spectrum of anomalous dimensions
of SYM operators, which can be extracted from the Thermodynamic
Bethe Ansatz of Refs.\mycite{Gromov:2009tv, Gromov:2009bc,
Gromov:2011cx}, there is still a lot of work to do to get an
acceptable comprehension of the three point functions. Yet
correlators in some specific sectors of the theory, $i.e.$ in
several well-defined limits, have been obtained in the form of
expansions in $\frac{1}{J}$ ($J$ being one of the $R$-charges),
$\frac{1}{N}$ and $\lambda' = \frac{\lambda}{J^2}$, i.e. as
finite-size, non-planarity and loop expansions correspondingly.
For some ``protected'' cases the results hold actually as exact
ones. The well-studied sectors of the theory are an important playground
for getting a more complete holographic understanding of the
three-point functions. On the field-theory side the computation of
the three-point functions is available for small coupling and
follows the planar perturbation theory.  On the string theory side
two different approaches are feasible. One is valid mostly for
strong coupling and is based upon the semiclassical methods, which
amounts to a calculation of a world-sheet with three singularities
that is equivalent to a three-point correlator in the field
theory. The other string-theoretical method is the string field
vertex construction, which on $AdS_5\times S^5$ is only known in
the pp-wave limit. An important feature of all the three
calculations is that there exists a limit in which all of them
{\it may} be valid, namely, in the so called Frolov-Tseytlin limit
of small $\lambda' =
\frac{\lambda}{J^2}$~\cite{Frolov:2003xy,Kruczenski:2003gt}.
Expansion in $\lambda'$ resembles a weak coupling expansion in
$\lambda$, yet is not identical to it, since while $\lambda'$ is
taken to be small, $\lambda$ can be either  small or large. The
Frolov-Tseytlin limit was originally conceived as a bridge between
weak and strong coupling for the non-protected operators in the
spectral problem. For the string states with angular momentum $J$
on $S^5$, the energies can be expanded in a limit of large $J$
around a BPS solution with $\lambda' = \frac{\lambda}{J^2}$ fixed.
This expansion can then be compared in the Frolov-Tseytlin limit
to the loop expansion on the gauge theory side. The energies match
the anomalous dimensions of the corresponding operators on the
gauge theory side up to and including the second order in the
expansion parameter, $i.e.$ two-loops on the gauge theory side,
but the matching breaks down at three-loops\mycite{Callan:2003xr}.
In\mycite{Harmark:2008gm} it is shown that the one-loop match is a
consequence of the suppression of quantum corrections to the
string near the BPS point, allowing a regime where the classical
action of the string is large even if a weak coupling expansion in
$\lambda$ is considered.

The first results for the three-point functions have shown that
the leading order calculations in string theory and field theory
do coincide, and this is already a non-trivial
statement~\cite{Escobedo:2010xs,Georgiou:2011qk}. However there
are some cases where the gauge theory and string theory results
have structure similarities but do not match perfectly even at the
leading order~\cite{Bissi:2011dc}. This can be presumably
interpreted as the inability of the basis chosen to describe gauge
theory operators to interpolate between weak and strong coupling.

Thus the aim of this work is to provide extra evidence  for the
gauge/string theory comparison in three point functions, using
operators for which the gauge and string identifications is very
well established~\cite{Berenstein:2002jq}. We perform in fact the
analysis of the BMN~\cite{Berenstein:2002jq} correlators with all
three momenta non-zero (the so-called fully dynamical correlators)
in the Frolov-Tseytlin limit. To the best of our knowledge, our
work is the first where this analysis is carried out for the
operators with all three momenta being non-zero.

To which of the classes -- heavy, light or intermediate -- do our
operators belong to ? Since $\Delta-J\sim \sqrt{1+\lambda' n^2}$,
at a fixed $\lambda'$ these operators represent an interesting
example of operators already heavy but still without an adequate
semiclassical description: taking $\lambda$ large, the anomalous
dimension $\Delta\sim J\equiv \sqrt{\lambda/\lambda'}$ can be made
scale as $\sqrt{\lambda}$. Thus we claim that at a fixed
$\lambda'$ our BMN operators are rather large. This will
eventually, as we hope, provide a solid ground to compare the
correlator of the (field-theory/pp-wave string-theory) BMN
operators/string states with a semiclassical correlator of giant
magnons, the latter being the ``heaviest'' objects available in
all possible senses of the definition.

It has recently been observed in\mycite{Bissi:2011ha} that there
is a discrepancy in the next-order $\lambda'$ expansion for
three-point correlators. The reason for this discrepancy is not
yet known, it might be due to the subtleties in the computation on
the gauge theory side.

For example, an apparent mismatch observed in an early stage of
three-point correlator studies\mycite{Kristjansen:2002bb} was
successfully resolved by finding a next-order in $1/N$ correction
to the operator-state identification rule -- a mixing of
single-trace with double trace operators was detected, since the
single-trace operators happened not to be the exact matches for
the string states.

In any case, understanding the cause for this mismatch is of
direct interest now. In doing so, extra evidence from other states
and sectors of the theory is of primary importance, since it can
possibly help us to distinguish between different causes: state
mixing, wrongly interpreted limits or, much less likely, some fundamental problems
with the duality conjecture.

The work is structured as follows: in
Section \ref{ft} we perform a field-theoretical calculation
of the correlator, and in Section \ref{st} we compute the same
correlator from string theory via the Dobashi-Yoneya 3-string vertex using the
asymptotic Neumann matrices in the pp-wave limit and make sure the
two results do agree. In the final Section \ref{cl} we comment on
the agreement between the two calculations and suggest
possible future directions of research in (dis)establishing the equivalence.

\section{Correlators of BMN operators}

Three-point correlators can be classified by the weights of the operators involved, these can be light (L), intermediate and heavy (H). By
definition, heavy state anomalous dimensions scale as
$\sqrt{\lambda}$
\beq\Delta_{heavy}\sim \sqrt{\lambda},\eeq
``intermediate'' states scale as
\beq\Delta_{intermediate}\sim \lambda^{\frac{1}{4}},\eeq
and the light states have
\beq\Delta_{light}\sim 1.\eeq
The three-point correlators are then classified in the simplest
approximation into LLL, LLH, LHH and HHH combinations. We can say
we know almost everything about the LLL; correlators from quite
some time~\cite{Lee:1998bxa,Freedman:1998tz,Arutyunov:1999en,Arutyunov:2000ku,
Bastianelli:2000mk,Heslop:2001gp,D'Hoker:2001bq,
Kristjansen:2002bb,Bianchi:2003ug,
Okuyama:2004bd,Georgiou:2009tp}
the HLL correlators are a bit exotic, they are mostly known from
the gauge theory side~\cite{Gromov:2011jh}; the HHL starting
from~\cite{Zarembo:2010rr,Costa:2010rz} have recently been and
continue to be an object of thorough research on both
integrability/field theory and string theory sides; there have
been some very promising attempts to construct also HHH
correlators both from
string~\cite{Dobashi:2002ar,Janik:2011bd,Buchbinder:2011jr,Kazama:2011cp}
and gauge theory
sides~\cite{Escobedo:2010xs,Foda:2011rr,Gromov:2012vu,Serban:2012dr,Kostov:2012jr}.

The object of our novel analysis are the fully dynamical
correlators of three BMN operators. They take an intermediate
position between the heavy and the light operators, since on the
one hand they do not possess a proper semiclassical description,
on the other hand at constant $\lambda'$ they scale as heavy
operators. Thus,  for large and small $\lambda$ they make a
perfect bridge towards the yet undisclosed domain of the HHH
correlators made of three giant magnons. For some reasons there is
a gap in the literature for BMN state correlators. Namely, the
results for the correlators of two BMN with one BPS are abundant,
whereas three BMN with three non-zero momenta have not been
calculated either on the gauge theory side (from the $1/N$
expansion of Feynman diagrams\mycite{Kristjansen:2002bb}), or on
string theory side using Neumann matrices provided
by\mycite{He:2002zu}. These are the calculations
done in the sections \ref{ft} and \ref{st} respectively.

There are however already some very interesting results on BMN
correlators. The topic was started from the string-theoretic point
of view
in\mycite{Spradlin:2002rv,Dobashi:2002ar,Pearson:2002zs,Pankiewicz:2002tg,He:2002zu,Pankiewicz:2003ap,DiVecchia:2003yp,
Dobashi:2004ka,Shimada:2004sw,Grignani:2005yv,Grignani:2006en} and from field
theory
in\mycite{Kristjansen:2002bb,Constable:2002hw,Beisert:2002bb}. The
three-point functions for two dynamical BMN and one static
(zero-momentum) operators on the field theory side up to first
order in $\lambda'$ were calculated in\mycite{Chu:2002pd}. Full
agreement with string theory has been found. An
``intermediate-intermediate-intermediate'' correlator of BMN
vacuum, BMN fermion-and-scalar excitation, BMN
fermion-and-scalar-and-an-R-charge excitation was calculated by
Dobashi in\mycite{Dobashi:2006fu} who pointed out the equality
between string and gauge theory results.

\subsection{\label{ft} BMN correlators from field theory}

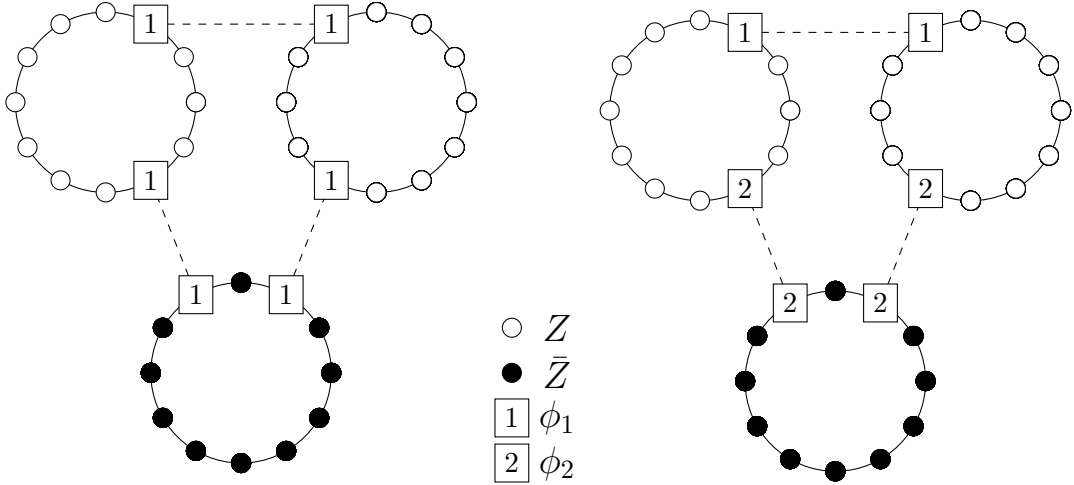
\begin{figure}
\begin{minipage}{0.5\textwidth}
\begin{tikzpicture}[scale=0.6]
\draw (0,0) circle (2cm);
\draw (6,0) circle (2cm);
\draw (3,-6) circle (2cm);
\foreach \x in {0,30,...,360}{
\draw[fill=white] (\x:2cm) circle(6pt);
\draw[fill=white] (8,0) circle(6pt);
\draw[fill=white] (7.73205,1) circle(6pt);
\draw[fill=white] (7,1.73205) circle(6pt);
\draw[fill=white] (6,2) circle(6pt);
\draw[fill=white] (8,0) circle(6pt);
\draw[fill=white] (7.73205,-1) circle(6pt);
\draw[fill=white] (7,-1.73205) circle(6pt);
\draw[fill=white] (6,-2) circle(6pt);
\draw[fill=white] (4,0) circle(6pt);
\draw[fill=white] (4.26795,1) circle(6pt);
\draw[fill=white] (5,1.73205) circle(6pt);
\draw[fill=white] (6,2) circle(6pt);
\draw[fill=white] (4,0) circle(6pt);
\draw[fill=white] (4.26795,-1) circle(6pt);
\draw[fill=white] (5,-1.73205) circle(6pt);
\draw[fill=white] (6,-2) circle(6pt);
\draw[fill=black] (5,-6) circle(6pt);
\draw[fill=black] (4.73205,-5) circle(6pt);
\draw[fill=black] (4,-4.26795) circle(6pt);
\draw[fill=black] (3,-4) circle(6pt);
\draw[fill=black] (1.26795,-5) circle(6pt);
\draw[fill=black] (2,-4.26795) circle(6pt);
\draw[fill=black] (4,-4.26795) circle(6pt);
\draw[fill=black] (1.26795,-7) circle(6pt);
\draw[fill=black] (2,-7.73205) circle(6pt);
\draw[fill=black] (3,-8) circle(6pt);
\draw[fill=black] (1,-6) circle(6pt);
\draw[fill=black] (4.73205,-7) circle(6pt);
\draw[fill=black] (4,-7.73205) circle(6pt);
\draw[fill=black] (3,-8) circle(6pt);
\draw[fill=black] (1,-6) circle(6pt);
\draw[fill=black] (1.26795,-7) circle(6pt);
}
\draw[dashed] (1,1.73205) -- (5,1.73205);
\draw[dashed] (5,-1.73205) -- (4,-4.26795);
\draw[dashed] (1,-1.73205) -- (2,-4.26795);
\path ( 1,1.73205) node [shape=rectangle,fill=white,draw] {1};
\path ( 5,1.73205) node [shape=rectangle,fill=white,draw] {1};
\path ( 5,-1.73205) node [shape=rectangle,fill=white,draw] {1};
\path ( 4,-4.26795) node [shape=rectangle,fill=white,draw] {1};
\path ( 1,-1.73205) node [shape=rectangle,fill=white,draw] {1};
\path (2,-4.26795) node [shape=rectangle,fill=white,draw] {1};
\draw[fill=white] (9,-5.) circle(6pt);
\draw (10,-5.) node(x)  {\Large $Z$};
\draw[fill=black] (9,-6.) circle(6pt);
\draw (10,-6.) node(x)  {\Large $\bar{Z}$};
\path ( 9,-7) node [shape=rectangle,fill=white,draw] {1};
\draw (10,-7) node(x)  {\Large $\phi_1$};
\path ( 9,-8) node [shape=rectangle,fill=white,draw] {2};
\draw (10,-8) node(x)  {\Large $\phi_2$};
\end{tikzpicture}
\end{minipage}
\begin{minipage}{0.49\textwidth}
\begin{tikzpicture}[scale=0.6]
\draw (0,0) circle (2cm);
\draw (6,0) circle (2cm);
\draw (3,-6) circle (2cm);
\foreach \x in {0,30,...,360}{
\draw[fill=white] (\x:2cm) circle(6pt);
\draw[fill=white] (8,0) circle(6pt);
\draw[fill=white] (7.73205,1) circle(6pt);
\draw[fill=white] (7,1.73205) circle(6pt);
\draw[fill=white] (6,2) circle(6pt);
\draw[fill=white] (8,0) circle(6pt);
\draw[fill=white] (7.73205,-1) circle(6pt);
\draw[fill=white] (7,-1.73205) circle(6pt);
\draw[fill=white] (6,-2) circle(6pt);
\draw[fill=white] (4,0) circle(6pt);
\draw[fill=white] (4.26795,1) circle(6pt);
\draw[fill=white] (5,1.73205) circle(6pt);
\draw[fill=white] (6,2) circle(6pt);
\draw[fill=white] (4,0) circle(6pt);
\draw[fill=white] (4.26795,-1) circle(6pt);
\draw[fill=white] (5,-1.73205) circle(6pt);
\draw[fill=white] (6,-2) circle(6pt);
\draw[fill=black] (5,-6) circle(6pt);
\draw[fill=black] (4.73205,-5) circle(6pt);
\draw[fill=black] (4,-4.26795) circle(6pt);
\draw[fill=black] (3,-4) circle(6pt);
\draw[fill=black] (1.26795,-5) circle(6pt);
\draw[fill=black] (2,-4.26795) circle(6pt);
\draw[fill=black] (4,-4.26795) circle(6pt);
\draw[fill=black] (1.26795,-7) circle(6pt);
\draw[fill=black] (2,-7.73205) circle(6pt);
\draw[fill=black] (3,-8) circle(6pt);
\draw[fill=black] (1,-6) circle(6pt);
\draw[fill=black] (4.73205,-7) circle(6pt);
\draw[fill=black] (4,-7.73205) circle(6pt);
\draw[fill=black] (3,-8) circle(6pt);
\draw[fill=black] (1,-6) circle(6pt);
\draw[fill=black] (1.26795,-7) circle(6pt);
}
\draw[dashed] (1,1.73205) -- (5,1.73205);
\draw[dashed] (5,-1.73205) -- (4,-4.26795);
\draw[dashed] (1,-1.73205) -- (2,-4.26795);
\path ( 1,1.73205) node [shape=rectangle,fill=white,draw] {1};
\path ( 5,1.73205) node [shape=rectangle,fill=white,draw] {1};
\path ( 5,-1.73205) node [shape=rectangle,fill=white,draw] {2};
\path ( 4,-4.26795) node [shape=rectangle,fill=white,draw] {2};
\path ( 1,-1.73205) node [shape=rectangle,fill=white,draw] {2};
\path (2,-4.26795) node [shape=rectangle,fill=white,draw] {2};
\end{tikzpicture}
\end{minipage}
\caption{\label{corr-3-point} Connected diagrams
contributing to the three-point function.}
\end{figure}

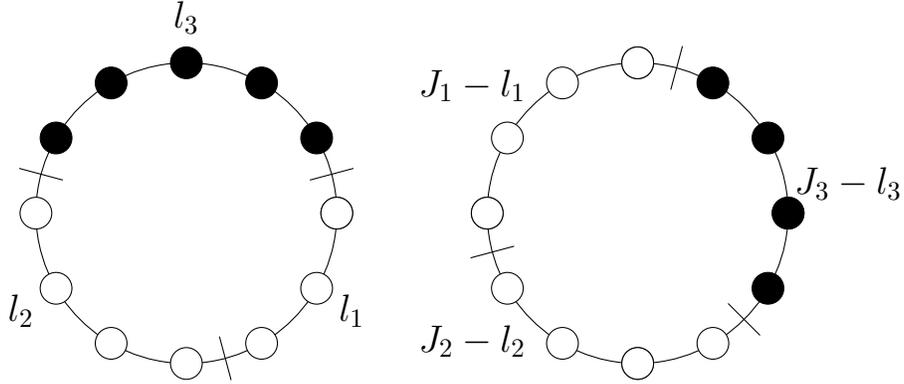
\begin{figure}
\begin{center}
\begin{tikzpicture}[scale=1]
\draw (0,0) circle (2cm);
\draw (6,0) circle (2cm);
\draw[fill=black] (0:2cm) circle(6pt);
\draw[fill=black] (30:2cm) circle(6pt);
\draw[fill=black] (60:2cm) circle(6pt);
\draw[fill=black] (90:2cm) circle(6pt);
\draw[fill=black] (120:2cm) circle(6pt);
\draw[fill=black] (150:2cm) circle(6pt);
\draw[fill=white] (180:2cm) circle(6pt);
\draw[fill=white] (210:2cm) circle(6pt);
\draw[fill=white] (240:2cm) circle(6pt);
\draw[fill=white] (270:2cm) circle(6pt);
\draw[fill=white] (300:2cm) circle(6pt);
\draw[fill=white] (330:2cm) circle(6pt);
\draw[fill=white] (360:2cm) circle(6pt);
\draw[fill=black] (8,0) circle(6pt);
\draw[fill=black] (7.73205,1) circle(6pt);
\draw[fill=black] (7,1.73205) circle(6pt);
\draw[fill=black] (6,2) circle(6pt);
\draw[fill=black] (8,0) circle(6pt);
\draw[fill=black] (7.73205,-1) circle(6pt);
\draw[fill=white] (7,-1.73205) circle(6pt);
\draw[fill=white] (6,-2) circle(6pt);
\draw[fill=white] (4,0) circle(6pt);
\draw[fill=white] (4.26795,1) circle(6pt);
\draw[fill=white] (5,1.73205) circle(6pt);
\draw[fill=white] (6,2) circle(6pt);
\draw[fill=white] (4,0) circle(6pt);
\draw[fill=white] (4.26795,-1) circle(6pt);
\draw[fill=white] (5,-1.73205) circle(6pt);
\draw[fill=white] (6,-2) circle(6pt);

\draw  (1.64207,0.439992) to (2.22163,0.595284);
\draw (-2.22163,0.595284) to (-1.64207,0.439992);
\draw (0.439992,-1.64207) to (0.595284,-2.22163);

\draw  (6.43999,1.64207) to (6.59528,2.22163);
\draw (3.77837,-0.595284) to (4.35793,-0.439992);
\draw (7.20208,-1.20208) to (7.62635,-1.62635);



\node[] at (0,2.6) {\Large $l_3$};
\node[] at (2.2,-1.3) {\Large$l_1$};
\node[] at (-2.2,-1.3) {\Large$l_2$};
\node[] at (3.8,1.7) {\Large$J_1-l_1$};
\node[] at (3.8,-1.7) {\Large$J_2-l_2$};
\node[] at (8.8,0.4) {\Large$J_3-l_3$};
\end{tikzpicture}
\caption{\label{twocircles}Two decoupled single-trace vacuum diagrams, $l_3=l_1+l_2$. The $l_1\ (J_1-l_1)$ and $l_2\ (J_2-l_2)$ $Z$-operators are divided to recall from which operators they originally came from.}
\end{center}
\end{figure}

Here we consider the computation of the three-point correlation function of BMN operators with non-zero momentum
in the weak coupling leading order in $1/N$, the  leading and the
next-to-leading order in the $1/J$ expansion. The operators we are interested in are single trace scalar operators defined as \beq\mathcal{O}_{ij,n}^{J}=\frac{1}{\sqrt{J
N^{J+2}}}\sum^J_{l=0}\tr \left(\phi_i Z^l\phi_j
Z^{J-l}\right)\psi_{n,l},\eeq which belong to the three
irreducible representations of $SO(4)$
\beq \bf 4\otimes 4=1+6+9,\eeq
where $\mathbf 1$ is the trace (T), $\mathbf 6 $ is the
antisymmetric (A), $\mathbf 9$ is the symmetric traceless
representation (S). The orthonormal basis therefore is
\beq \begin{array}{l}
A_{ij}=\frac{1}{\sqrt{2}}\left(\mathcal{O}_{ij}
-\mathcal{O}_{ji}\right),\\ \\
S_{ii}=\frac{2}{\sqrt{3}}\left(\mathcal{O}_{ii}-
\frac{1}{4}\sum_{i'}\mathcal{O}_{i'i'}\right),\\ \\
\tilde{S}_{ij}=\frac{1}{\sqrt{2}}\left(\mathcal{O}_{ij}
+\mathcal{O}_{ji}\right),\\ \\
T=\frac{1}{2}\sum_{i'}\mathcal{O}_{i'i'}.
\end{array}
\eeq
To simplify the notation we omit the momentum indices $n_i$. We shall be interested
in the leading-order $1/N$ behavior solely, therefore, we do not
take into account the mixing of single-trace with double trace
operators that takes place at the next-order. The
wave-functions for different representations are
\beq\begin{array}{l} \psi^S_{n,l}=\cos\frac{(2l+1)\pi n}{J+1},\\
\\
\psi^A_{n,l}=\sin\frac{2(l+1)\pi n}{J+2},\\ \\
\psi^T_{n,l}=\cos\frac{(2l+3)\pi n}{J+3}.\end{array} \eeq
We consider the correlation function of three BMN fully dynamical operators which is given by
\beq C_{i_1 j_1, n_1;i_2 j_2, n_2;i_3 j_3, n_3}^{J_1 J_2
J}=\langle \mathcal{O}_{i_1j_1,n_1}^{J_1}
\mathcal{O}_{i_2j_2,n_2}^{J_2}\bar{\mathcal{O}}_{i_3j_3,n_3}^{J}\rangle,\eeq
where no extra overall normalization has been introduced since the
operators $\mathcal{O}_{ij,n}^{J}$ are already unity-normalized. We denote the correlator of
three BMN operators taken with non-zero momentum as ``fully dynamical'', unlike e.g. the
three-point correlator mentioned in\mycite{Kristjansen:2002bb},
which, having one vanishing momentum, can be denoted
as ``partially dynamical'',  being  a correlator of two
BMN and one chiral primary. As already mentioned, the obvious
generalization to a fully dynamical correlator has not yet been considered in the literature. The R-charges $J$ have to be
conserved, therefore $J=J_1+J_2$. For convenience below we shall
use the notation
\beq J_1=rJ,\,\,\,  J_2=(1-r)J,\eeq
where the parameter $r$ is understood as a finite fixed quantity, $0\le r\le 1$,
and we consider the large $J$ limit. We are interested
only in the contribution to the 3-point correlation function coming from the connected diagrams. An example of such a diagram is given in
\myfigref{corr-3-point}.
%
%
This diagram is evaluated as prescribed
in\mycite{Kristjansen:2002bb}. One first contracts the impurity operators and this leads to two decoupled
single-trace vacuum diagrams, as that shown in\myfigref{twocircles}, and its counterparts with respect to the transformation $l\to
J-l$. The diagram\myfigref{twocircles} corresponds to the quantity
\beq \left\langle\tr\left(Z^{l_1}Z^{l_2}\bar{Z}^{l_3}\right)
\tr\left({Z}^{J_1-l_1}{Z}^{J_2-l_2}\bar{Z}^{J-l_3}\right)
\right\rangle. \eeq
The $l_1 (J_1-l_1)$ and $l_2 (J_2-l_2)$ $Z$-operators are separated to recall from which operators they originally came from.
Since we work in the leading-order approximation in $1/N$, the
diagram is evaluated as disconnected and simply equals to
$N^{J+2}$.  Disconnectedness of this diagram imposes the condition $l_3=l_1+l_2$. There are 4 diagrams in total like those in\myfigref{twocircles}.
Let the diagram in\myfigref{twocircles} be equal to
$f(l_1,l_2,l_3)$. The full contribution to the correlator is
then
\beq C_{IJK}\sim
f(l_1,l_2,l_1+l_2)+f(l_1,l_2,J_1-l_1+l_2)+f(l_1,l_2,J_2+l_1-l_2)+
f(l_1,l_2,J-l_1-l_2).\eeq The answer for the correlator is given
by a convolution of the three corresponding wave-functions with
this expression.

\begin{figure}[h]
\begin{center}
\includegraphics[height =5cm, width=9cm]{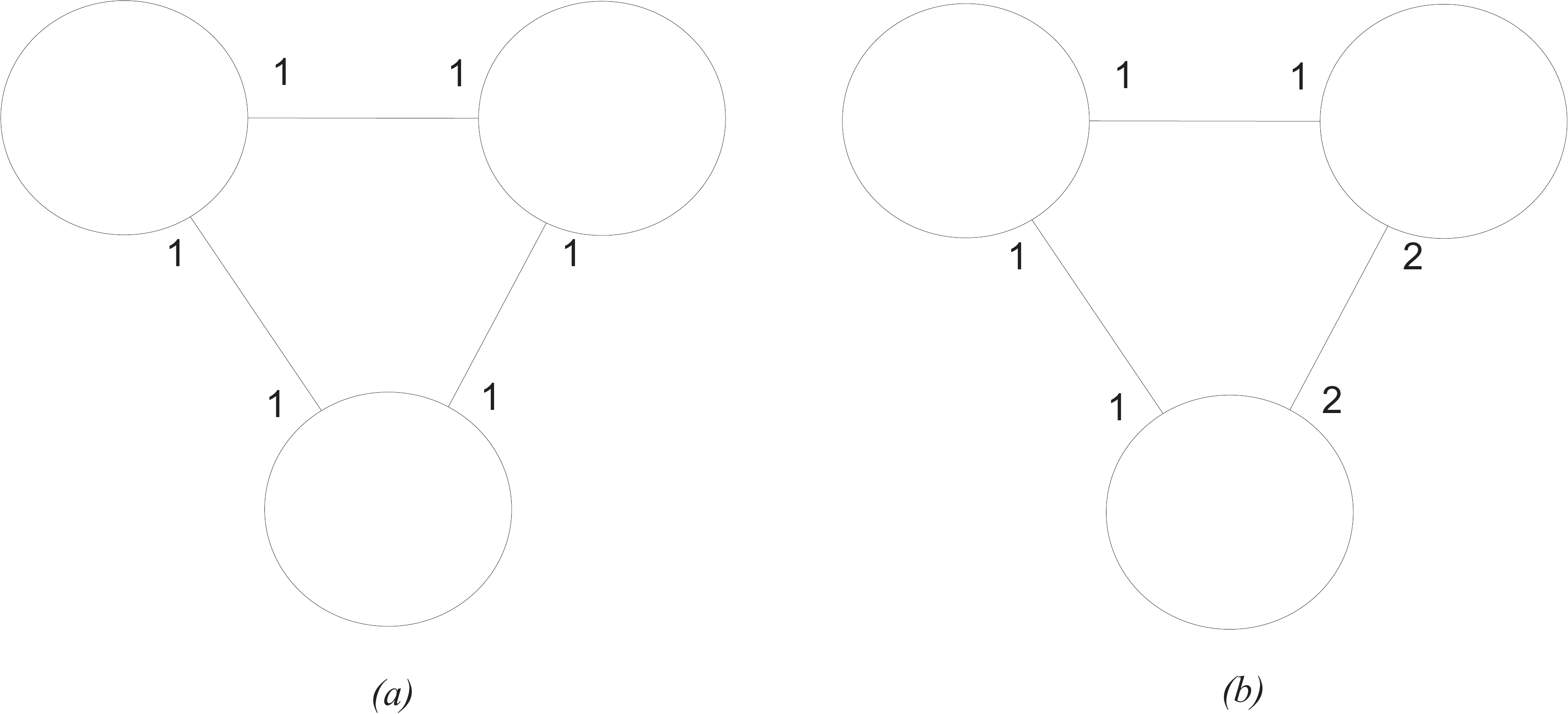}
\caption{\label{combin}Two possible types of
the $(i_1 j_1) (i_2 j_2) (i_3 j_3)$ indices that can contribute.}
\end{center}
\end{figure}

Due to $SO(4)$ charge conservation there are two possible types of
the $(i_1 j_1) (i_2 j_2) (i_3 j_3)$ indices that can contribute,
as shown in\myfigref{combin}: $(ii),(ii),(ii)$ (Fig. (a)) and
$(ii) (ij) (ij)$ (Fig. (b)).
\vskip .2 cm

There are four $SO(4)$ irrep structures corresponding
to\myfigref{combin} (a): $SSS,SST,STT,TTT$ and six $SO(4)$ irrep
structures corresponding to\myfigref{combin} (b):
$S\tilde{S}\tilde{S},S\tilde{S}A,SAA,T\tilde{S}
\tilde{S},T\tilde{S}A,TAA$. The irrep combinatorics is
supplemented by permutations of $n_1,n_2,n_3$. Since
$n_1\leftrightarrow n_2,J_1\leftrightarrow J_2$ is a trivial
symmetry of the three-point function under consideration, the
total amount of combinations can be handled, and we do not show
the correlators that differ only by a permutation of the two first
operators. In Table \ref{tab1}  we list all of the remaining structures. The first letter refers to the wavefunction with
momentum quantum number $n_1$, the second with $n_2$, the third
with $n_3$.
\begin{table}[h!]\caption{\label{tab1}Possible
configuration of the three-point functions.}\beq\begin{array}{l}
SSS\\
SST,TSS\\
STT,TTS\\
TTT\\
S\tilde{S}\tilde{S}\\
S\tilde{S}A,SA\tilde{S},\tilde{S}AS\\
SAA,AAS\\
T\tilde{S}\tilde{S}\\  T\tilde{S}A,TA\tilde{S},\tilde{S}AT\\
TAA,AAT
\end{array}\eeq
\end{table}
One immediately sees that all correlators where the third operator
is an antisymmetric one vanish due to the property of the
antisymmetric wave functions. Also the $\tilde{S}$ and $S$ states
after the internal $ij$ lines have been contracted differ only by
a constant multiplicative factor. We can summarize the
table\myref{tab1} in terms of few simpler objects
using the orthonormal basis defined above:
\beq\label{tab2}\begin{array}{ll}
SSS=\frac{3}{8} n_s^3c_{SSS},\\ \\
SST=\frac{3}{4}n_s^2n_tc_{SST},&TSS=\frac{3}{4}n_s^2n_tc_{TSS},\\
\\
STT=0,&TTS=0,\\ \\
TTT=4 n_t^3 c_{TTT},\end{array}\eeq

\begin{equation*}\begin{array}{ll}
S\tilde{S}\tilde{S}=\frac{1}{2} n_s n_{\tilde{s}}^2 c_{SSS},\\ \\
AS\tilde{S}=\frac{1}{2} n_a n_s n_{\tilde{s}} c_{ASS},&
A\tilde{S}S=\frac{1}{2}  n_a n_s n_{\tilde{s}} c_{ASS},\\ \\
AAS=\frac{1}{2} n_a^2 n_s c_{AAS},\\ \\
T\tilde{S}\tilde{S}=4 n_t n_{\tilde{s}}^2 c_{TSS},\\
\\ TA\tilde{S}=4 n_t n_a n_{\tilde{s}} c_{TAS},&
\tilde{S}AT=4 n_t n_a n_{\tilde{s}} c_{SAT},\\ \\AAT=4 n_a^2 n_t
c_{AAT}. \end{array}\end{equation*}
Here the norms are $n_a=n_{\tilde{s}}=\frac{1}{\sqrt{2}}$,
$n_s=\frac{2}{\sqrt{3}}$, $n_t=2$. The coefficients $c_{IJK}$ are
defined as the correlators of the operators:
$\mathcal{O}_{11}^I$ (where $I=S,T$) and $\mathcal{O}_{12}^I$
(where $I=\tilde{S},A$). The order of letters reflects the
cardinal numbers of the momenta $n_1,n_2,n_3$.
Calculating the $c_{IJK}$ directly we find that there are only four non-zero contributions: $SSS,SST,TTS$ and $TTT$. The correlators
are known to us in the leading $1/N$ order and up to the
subleading $1/J$ order. Defining
\beq c_{IJK}=\frac{J^{1/2}}{N}\left(c_{IJK}^0+
\frac{1}{J}c_{IJK}^1\right), \eeq
after the calculation we see that \beq\label{res1}
c^0_{SSS}=c^0_{SST}=c^0_{STT}=c^0_{TTT}=\frac{-4n_3^2
r^{3/2}(1-r)^{3/2} \sin(\pi n_3
r)^2}{\pi^2(n_2^2-(1-r)^2n_3^2)(n_1^2-r^2n_3^2)},\eeq
and for the subleading part one gets the structures
\beq\label{res}
c^1_{IJK}=\frac{1}{\pi^2(n_2^2-(1-r)^2n_3^2)^2(n_1^2-r^2n_3^2)^2}
\bar{c}^1_{IJK}, \eeq
which for $|n1|=r|n3|$ or  $|n2|=(1-r)|n3|$ is singular  since it
is multiplied by the following regular numerators
\beq
\begin{array}{rcl}
\displaystyle  \bar{c}^1_{SSS}&=&4 n_3^2 (r-1) r \sin \left(\pi
n_3 r\right) \left(\pi n_3 r \left(2 r^2-3 r+1\right) \left(n_3^2
(r-1)^2-n_2^2\right) \left(n_3^2 r^2-n_1^2\right) \cos \left(\pi
n_3 r\right)+ \right.\\ \\  &&\displaystyle \left. \,\, +2
\left(n_1^2 \left(n_2^2 \left(r^2-r+1\right)+n_3^2
(r-1)^3\right)-n_3^2 r^3 \left(n_3^2 (r-1)^3+n_2^2\right)\right)
\sin \left(\pi  n_3 r\right)\right),\end{array} \eeq
\beq
\begin{array}{rcl}
\displaystyle \bar{c}^1_{SST}&=&4 n_3^2 (r-1) r \sin \left(\pi n_3
r\right) \left(3 \pi  n_3 r \left(2 r^2-3 r+1\right) \left(n_3^2
(r-1)^2-n_2^2\right) \left(n_3^2 r^2-n_1^2\right) \cos \left(\pi
n_3 r\right)+\right. \\ \\ &&\displaystyle \left. \,\, +2
\left(n_1^2 \left(n_2^2 \left(3 r^2-3 r+1\right)+n_3^2
(r-1)^3\right)-n_3^2 r^3 \left(3 n_3^2 (r-1)^3+n_2^2\right)\right)
\sin \left(\pi n_3 r\right)\right),\end{array} \eeq
\beq
\begin{array}{rcl}
\displaystyle \bar{c}^1_{STT}&=&-4 n_3 (r-1) \sin \left(\pi  n_3
r\right) \left(\pi (r-1) \left(n_2^2-n_3^2 (r-1)^2\right)
\left(n_3^4 r^4 (2 r-1)-\right.\right.\\ \\ && \displaystyle
\left.\left. \,\, -n_3^2 n_1^2 r^2 (2 r+1)+2 n_1^4\right) \cos
\left(\pi n_3 r\right)+\right. \\ \\ && \displaystyle \left.+2 n_3
r \left(n_3^2 r^3 \left(n_3^2 (r-1)^3+n_2^2\right)-n_1^2
\left(n_2^2 \left(r^2-3 r+3\right)+3 n_3^2 (r-1)^3\right)\right)
\sin \left(\pi  n_3 r\right)\right),\end{array} \eeq
\beq
\begin{array}{rcl}
&&\displaystyle \bar{c}^1_{TTT}=4 n_3 \sin \left(\pi n_3 r\right)
\left(6 n_3 (r-1) r \left(n_1^2 \left(n_2^2
\left(r^2-r+1\right)+n_3^2 (r-1)^3\right)-\right.\right. \\ \\ &&
\displaystyle \left.\left. - n_3^2 r^3 \left(n_3^2
(r-1)^3+n_2^2\right)\right) \sin \left(\pi  n_3 r\right)-\right.\\
\\ && \displaystyle \left.- \pi \left(n_2^2-n_3^2 (r-1)^2\right)
\left(n_3^2 r^2-n_1^2\right) \left(r^2 \left(3 n_3^2 (r-1)^2 (2
r-1)+2 n_2^2\right)-2 n_1^2 (r-1)^2\right) \cos \left(\pi  n_3
r\right)\right).\end{array} \eeq
Note that all four structures are different from each
other in the subleading order. To make some sense from these
illegible expressions let us expand
for small momenta, $n_i\to 0$. This will correspond to the near-BMN limit. We
get then
\beq
\begin{array}{l} \\ \displaystyle
\bar{c}^1_{SSS}=4 \pi ^2 n_1^2 n_2^2 n_3^4 r^3 \left(4 r^3-9 r^2+8
r-3\right),\\ \\ \displaystyle\bar{c}^1_{SST}=4 \pi ^2 n_1^2 n_2^2
n_3^4 r^3 \left(12 r^3-27 r^2+20 r-5\right),\\ \\ \displaystyle
\bar{c}^1_{STT}=4 \pi ^2
n_1^2 n_2^2 n_3^4 r^3 \left(4 r^3-11 r^2+12 r-5\right),\\
\\ \displaystyle \bar{c}^1_{TTT}=12 \pi ^2 n_1^2 n_2^2 n_3^4 r^3 \left(4 r^3-9
r^2+8 r-3\right).
\end{array}
\eeq
The leading order part of our results resembles (perhaps not
surprisingly) the expressions
obtained in\mycite{Klose:2011rm}.
Now one could consider  comparing these expressions to
semiclassical calculations. They must not necessarily coincide,
since the above calculation has been performed at weak coupling.
Therefore such a comparison will be highly non-trivial. The
closest objects on the strong coupling side to our BMN operators
are the giant magnons. They require a full two-dimensional
analysis of the worldsheet configurations, unlike the long BPS
operators considered by~\cite{Klose:2011rm,Buchbinder:2011jr} that
effectively reduced the classical worldsheet to a combination of
geodesics. We postpone this truly semiclassical analysis to a
successive work, and now proceed in Section~\ref{st} to a doable
yet nontrivial comparison with the matrix elements of the string
interaction Hamiltonian 3-vertex in the pp-wave limit.

\subsection{\label{st}The three-point BMN correlator from string theory}
About a decade ago a very advanced technique was developed for
calculating the light-cone  string-theory three-point matrix
elements of the interaction Hamiltonian. The general idea of the
calculation is that a matrix element $H_{IJK}\equiv \langle
IJK|H|0\rangle$ is obtained from  the construction
\beq H_{IJK}= \langle IJK| P|V \rangle,\eeq
where the exponential factor is
\beq |V\rangle
=e^{\frac{1}{2}\sum_{a,b,i,j}N^{a,b}_{i,j}a^\dagger_{ai}
a^\dagger_{bj}}|0\rangle~,\eeq
the matrices $N^{a,b}_{i,j}$ are the Neumann matrices,
the indices $a,b$ running through 1 to 3 and corresponding to the
states $IJK$, the indices $i,j$ corresponding to the oscillator
modes.
The most advanced three string vertex in the pp-wave limit~\cite{Grignani:2005yv,Grignani:2006en} was found by
Dobashi and Yoneya\mycite{Dobashi:2004ka} as a linear combinations with equal weight of the vertices proposed in~\cite{Spradlin:2002rv,DiVecchia:2003yp}. The
prefactor $P$ is organized as
\beq
\begin{array}{l}\displaystyle
P=\frac{\omega_{1,n_1}}{\mu r} \left(2+\alpha_{1,n_1}^\dagger
\alpha_{1,-n_1}\vphantom{\dagger}+\alpha_{1,-n_1}^\dagger
\alpha_{1,n_1}\vphantom{^\dagger}\right)
+\frac{\omega_{2,n_2}}{\mu (1-r)} \left(2+\alpha_{2,n_2}^\dagger
\alpha_{2,-n_2}\vphantom{^\dagger}+\alpha_{2,-n_2}^\dagger
\alpha_{2,n_2}\vphantom{^\dagger}\right) -\\ \\
\displaystyle-\frac{\omega_{3,n_3}}{\mu} \left
(2+a_{3,n_3}^\dagger
\alpha_{3,-n_3}\vphantom{^\dagger}+\alpha_{3,-n_3}^\dagger
\alpha_{3,n_3}\vphantom{^\dagger}\right).
\end{array}
\eeq
Here $\mu$ is the expansion parameter of the Penrose limit,
$\mu\sim \frac{1}{\sqrt{\lambda'}}$. The frequencies $\omega$ are
\beq
\begin{array}{l}\displaystyle
\omega_{1,n}=\sqrt{n^2+\mu^2 r^2},\\ \\ \displaystyle
\omega_{2,n}=\sqrt{n^2+\mu^2 (1-r)^2},\\ \\ \displaystyle
\omega_{3,n}=\sqrt{n^2+\mu^2}.
\end{array}
\eeq
We do not discuss here the fermionic contribution to the
prefactor, which caused a lot of dispute in the literature, where
at least three different types of vertices have been
compared\mycite{Grignani:2006en}. This discussion is so far
irrelevant to us since all our states are bosonic. The matrices
$N^{a,b}_{i,j}$ are taken by us from\mycite{He:2002zu,Grignani:2005yv,Grignani:2006en}. Their
behaviour for the positive and the negative values of the mode
numbers is essentially different. For the positive modes $m,n$ the
leading-order in $\mu$ (up to $\mathcal{O}(\mu^1)$) is
\beq N_{m,n}=\left(
\begin{array}{ccc}
\frac{(-1)^{m+n}}{2 \mu  \pi  r} &  -\frac{(-1)^m}{2 \mu  \pi
\sqrt{(1-r) r}} &  -\frac{2 (-1)^{m+n} n
r^{3/2} \sin (n \pi r)}{\pi  \left(n^2 r^2-m^2\right)} \\ \\

 -\frac{(-1)^n}{2 \mu  \pi  \sqrt{(1-r)r}}
&  -\frac{1}{2 \mu  \pi (r-1)} &  -\frac{2 (-1)^n n
(1-r)^{3/2} \sin (n \pi  r)}{\pi (m^2-n^2 (1-r)^2) } \\ \\

 -\frac{2(-1)^{m+n} m r^{3/2}  \sin (m \pi  r)}{\pi
\left(m^2 r^2-n^2\right)} & -\frac{2 (-1)^m m (1-r)^{3/2} \sin (m
\pi r)}{\pi  (n^2-m^2 (1-r)^2) } & 0
\end{array}
\right). \eeq
For negative modes $-m,-n$ the Neumann matrix becomes
\beq N_{-m,-n}=\left(
\begin{array}{ccc}
 0 &  0 &  -\frac{2 (-1)^{m+n} m \sqrt{r} \sin (n \pi  r)}{\pi
  \left(m^2-n^2 r^2\right)} \\ \\
 0 & 0 & -\frac{2 (-1)^n m (r-1)^{1/2}
  \sin (n \pi  r)}{\pi
  \left(n^2 (r-1)^2-m^2\right) } \\ \\
 -\frac{2 (-1)^{m+n} n \sqrt{r} \sin (m \pi  r)}{\pi
  \left(n^2-m^2 r^2\right)} &  -\frac{2 (-1)^m n
   (r-1)^{1/2} \sin (m \pi  r)}{\pi
   \left(m^2 (r-1)^2-n^2\right) } &
    \frac{2 (-1)^{m+n} \sin
   (m \pi  r) \sin (n \pi  r)}{\mu  \pi }
\end{array}
\right).\eeq
The idea behind  the comparison between the correlation function
and Hamiltonian matrix element is the conjecture
\beq\label{conje} \langle i H jk \rangle \sim \mu
\left(\Delta_i-\Delta_j-\Delta_k \right) C_{ijk}, \eeq
where the correlator $C_{ijk}$ is exactly what we have just
calculated in the previous section
\beq
C_{ijk}=\langle\bar{\mathcal{O}}_i\mathcal{O}_j\mathcal{O}_k\rangle.\eeq
It is supposed that the string states are identified in some
well-defined way with the single-trace gauge theory operators.
This is not really true\mycite{Beisert:2002bb}, due to mixing with
double-trace operators~\footnote{We specially thank Gordon
Semenoff for a discussion of this point.}, but we omit here this
discussion, since in the leading $\mu$ and $\lambda'$ order it is
irrelevant. In the next-leading order in $\mu$ the operator
redefinition will have to be taken into account.
When identifying gauge theory operators with the string theory
states we should also note the different oscillator bases used.
Namely, the natural spin chain/gauge theory creation operator is
given by $\alpha_n^\dagger$, whereas the natural string theory
operators are denoted by $a^\dagger$. The relation between them is
\beq \label{redef}\alpha_{n}=\frac{a_{|n|}
-i\mathrm{sign}(n)a_{-|n|}}{\sqrt{2}}.
\eeq
String theory states in the matrix element $H_{123}$ are defined
as $a^\dagger|0\rangle$. Field theory oscillators in $C_{123}$ are
defined as $\alpha^\dagger|0\rangle$.
With all normalizations taken into account, the
conjecture\myref{conje} will boil down to the following working
formula that is given by Dobashi and Yoneya~\cite{Dobashi:2004ka} and rewritten in our
notations at the leading-order in the large $\mu$ limit as (see their eq.
(2.2), (3.9))
\beq C_{123}=\frac{1}{2\mu}\frac{\sqrt{J_1 J_2 J}}{N}
\left(\frac{J}{4\pi \mu}\right)^{-1} H_{123}. \eeq
We have already taken into account here that
$\Delta_3-\Delta_2-\Delta_1=2$. The matrix element $H_{123}$ is
organized as
\beq H_{123}=\frac{1}{8} P N^3, \eeq
where we symbolically denote by $P$ the prefactor contribution, by
$N^3$ the exponential contribution; the $1/8$ factor comes from
the operator redefinition\myref{redef}. For simplicity we take the
case of three similar excitations, like the $SSS$ case in the
previous section. Doing the elementary algebra we get
\beq P=4,\eeq
and
\beq
\begin{array}{l}\displaystyle
N^3=-N^{12}_{n_1, n_2}N^{23}_{n_2, n_3}N^{31}_{n_3, n_1}
-N^{11}_{n_1, n_1}N^{23}_{n_2, n_3}N^{23}_{n_2, n_3}-\\ \\
\displaystyle + N^{12}_{-n_1, -n_2}N^{23}_{-n_2,
-n_3}N^{31}_{-n_3, -n_1} +N^{11}_{-n_1, -n_1}N^{23}_{-n_2,
-n_3}N^{23}_{-n_2, -n_3},
\end{array}
\eeq
where we have taken into account the combinatorial factor $48$ (8
possible choices of pairings $\times$ 6 permutations), canceled
with the factor $1/48$ coming from the exponent. Noticing that the
piece with $N^{11}$ exactly  corresponds to a disconnected
diagram, the connected sector, equivalent to the diagram
in\myfigref{combin} is simply given by
\beq
\begin{array}{l}\displaystyle
N^3_{connected}=-N^{12}_{n_1, n_2}N^{23}_{n_2, n_3}N^{31}_{n_3,
n_1} + N^{12}_{-n_1, -n_2}N^{23}_{-n_2, -n_3}N^{31}_{-n_3, -n_1} ,
\end{array}
\eeq
Gathering all the coefficients and expanding the product of
Neumann matrices, we get
\beq C_{123}=\frac{-4 r^{3/2} (1-r)^{3/2} \sin^2 \pi n y}{\pi^2
\left (n_2^2-(1-r)^2n_3^2\right)\left(n_1^2 -r^2 n_3^2\right)},
\eeq
fully agreeing to the leading-order of our correlator in the
previous section\myref{res1}. This agreement is the main result of
our work.

\section{Conclusions and outlook\label{cl}}
In this work we have demonstrated that the fully dynamical
correlator of three BMN states, each with two impurities and with
a non-zero momentum, as calculated field-theoretically with the
procedure of~\cite{Beisert:2002bb} completely agrees with the
string-theoretical calculation of the 3-string vertex matrix
elements, as proposed
in~\cite{Pankiewicz:2003ap,Pankiewicz:2002tg,He:2002zu,DiVecchia:2003yp,Dobashi:2004ka}.

In~\cite{Bissi:2011ha} a remarkable discrepancy at
the next order in $\lambda'$ was found
for three-point correlators from strings and from field theory. Our leading-order result establishes a
firm ground for the next order, a comparison which will be the next logical
step to be done.
Another extension that naturally follows from our work
would be to look at the finite-size corrections, as done for the
three-point correlators in~\cite{Bozhilov:2011qf,Arnaudov:2011ek}.
Finite-size corrections often do provide non-trivial tests of the
AdS/CFT correspondence, e.g. as for a different sector of
observables was done in our work\mycite{Astolfi:2011bg}.

While the two suggested further research directions  -- doing the
next-order in $\lambda'$ and $\frac{1}{J}$ are in principle feasible,
going along the path suggested in~\cite{Zarembo:2010rr,Buchbinder:2011jr,Klose:2011rm} and performing a true strongly-coupled
semiclassical calculation in order to move to the giant magnon end
of the asymptotic space, would be a true challenge and a step into
the {\it terra incognita} for our class of states. One would
certainly be interested in proceeding to the correlator of
giant-magnon-like heavy states with all momenta non-zero from our
``heavy'' (at fixed $\lambda'$)  BMN ones. To achieve that goal
several conceptual problems have to be solved, the most important
of them is how to match the three giant magnons
world-sheets\footnote{We specially thank Tristan McLoughlin for an
interesting correspondence on that point.}.

Such heavy-heavy-heavy correlators are certainly the most
mysterious and the least known beasts in the three-point bestiary.
Yet taking the existence of a smooth transition from giant magnons
to simply heavy magnons, and the miracles observed for the
correspondence between small $\lambda'$ and large $\lambda'$
expansions, our calculations would provide at least a starting
point for comparison of correlators of three giant magnons. In view
of the alleged universality of $\lambda'$ asymptotics for both
large and small $\lambda$, the asymptotic way to the semiclassic
regime seems to be feasible.

\section*{Acknowledgments}

We thank Agnese Bissi, Troels Harmark, Tristan McLoughlin, Marta
Orselli, Gordon Semenoff and Arkady Tseytlin for interesting and
stimulating discussions. This work was supported in part by the
MIUR-PRIN contract 2009-KHZKRX. The work of A.Z. is supported in
part by the Ministry of Education and Science of the Russian
Federation under contract 14.740.11.0081, NSh 3349.2012.2, the
RFBR grants 10-01-00836 and 10-02-01483.


\providecommand{\href}[2]{#2}\begingroup\raggedright\endgroup

\end{document}